\documentclass[letter]{aa} 
\usepackage{graphicx}
\usepackage{txfonts}
\begin{document} 

\title{Mass without radiation: Heavily
  obscured AGNs, the X--ray background, and the black hole mass density}

\author{
A.~Comastri\inst{1}
\and
R.~Gilli\inst{1}
\and
A.~Marconi\inst{2,3}
\and
G.~Risaliti\inst{3}
\and
M.~Salvati\inst{3}
}

\institute{
INAF -- Osservatorio Astronomico di Bologna, Via Ranzani 1, 40127 Bologna, Italy
\email{andrea.comastri@oabo.inaf.it}
\and
 Dipartimento di Fisica e Astronomia, Universit\`a di Firenze,
via Sansone 1, 50019 Sesto Fiorentino, Firenze, Italy
\and
INAF -- Osservatorio Astrofisico di Arcetri,  Largo E. Fermi 5,
50125 Firenze, Italy
}

\date{Received ??? Accepted ???}

\abstract{
 A recent revision of black hole scaling relations 
 indicate that  the local mass density in black holes should be 
  five times higher than previously determined values. 
 The local black hole mass density  is connected to the mean radiative
 efficiency of accretion through
 the time integral of the AGN volume density and a significant  increase in the local black hole mass density 
 would have interesting  consequences on AGN accretion properties and demography.
 One possibility to explain a large black hole mass density is that
 most of the black hole growth is via radiatively inefficient
 channels such as super Eddington accretion; however, this solution is not
 unique. Here we show how it is possible to accommodate a larger fraction of heavily buried, Compton-thick AGNs, 
 without violating the  limit imposed by the spectral energy density
 of the hard  X--ray and  mid--infrared backgrounds.}


\authorrunning{A. Comastri et al.}
\titlerunning{Mass without radiation}

\maketitle

\section{Introduction}

The strong correlations between the masses of the central super
massive  black holes (SMBH) 
and the global properties of their host spheroids such as
luminosities, dynamical masses and velocity dispersions, i.e. the 
scaling relations, can be used to convert the mass function of local
galaxies into  a black hole mass function and, by
integration, into a local SMBH mass density $\rho_{\bullet,loc}$.

A recent comprehensive analysis of black hole mass measurements and
scaling relations concluded that the canonical black-hole-to-bulge
mass ratio,
 $M_{SMBH}/M_{Bulge}$, shows a mass dependence and varies from 
0.1--0.2\% at    $M_{\rm bulge} \simeq 10^{9} M_{\odot}$ to 
$\sim 0.5$\%  at $M_{\rm bulge} = 10^{11} M_{\odot}$ (\citealt{GS13}, \citealt{KO13}). 
The revised normalization is a factor of 2 to 5
larger than previous estimates ranging from $\simeq$ 0.10\%
(\citealt{mf01}, \citealt{mclure02}, \citealt{Sani11}) to $\sim$
0.23\% \citep{marconi03}, therefore  resulting in an overall
increase in the normalization, which is dominated by massive bulges.

The mass function expected from black holes grown by
accretion during active phases  and the corresponding
integral over the cosmic time $\rho_{\bullet,acc}$, are closely related
to the AGN luminosity function  via a mass--to--luminosity
conversion factor. The  argument, originally proposed by \citet{Sol82},
can be formulated as a continuity
equation for the mass density of black holes
(e.g. \citealt{Marconi04}; \citealt{shankar04}; \citealt{merloni08}). 
Assuming a constant radiative efficiency  and 
neglecting mergers, the continuity equation can be integrated to obtain  

\begin{equation}
  \rho_{\bullet,acc}c^2 = U_T \frac{1- \epsilon}{\epsilon} 
  = \langle k_{BOL} \rangle U_X \frac{1- \epsilon}{\epsilon}  
\label{eq:grama1} 
,\end{equation}where $\epsilon$ is the radiative efficiency, $U_T$ is the bolometric, comoving energy density of AGNs
described by a bolometric luminosity function $\Phi(L,z)$, 

\begin{equation}
U_T = \int dz \frac{dt}{dz} \int L \Phi(L,z) dL
\label{eq:grama2} 
.\end{equation}

The comoving energy density  $U_T$  can be written as the product
of the X--ray energy density $U_X$ times an average bolometric correction
factor $\langle k_{BOL} \rangle$ to convert the X--ray
emissivity to  bolometric emissivity (e.g. \citealt{Beta12}).  

Equation 1  is widely used in AGN demographic studies to infer the efficiency of accretion processes 
by comparing the values obtained by the integration of the bolometric luminosity
function with  the local black hole mass density obtained from the scaling relations.
A good match between $\rho_{\bullet,loc}$ and $\rho_{\bullet,acc}$
is achieved for an accretion efficiency consistent with that 
expected from a standard \citet{SS73} accretion disk, broadly supporting the possibility that most of the
SMBH mass is efficiently accreted over the Hubble time (e.g. \citealt{fi99}). 
Using the X--ray background as an integral constraint and
 assuming that the bulk of the X--ray emissivity is at  $\langle
 z\rangle\simeq$ 2, 
\citet{Elvis02} infer a  lower limit on the accretion efficiency  of $\epsilon>$0.15,
  arguing that most SMBH are rapidly spinning. 
\par
Most recent estimates are in good agreement with
$\langle\epsilon\rangle\simeq$ 0.1.
For instance, \citet{Marconi04} used luminosity functions in the optical B-band
\citep{Boy00}, soft X--ray, 0.5--2 keV band \citep{taka00},
and hard X--ray,  2--10 keV band \citep{Ueda03}
transformed into  a bolometric luminosity function using a luminosity dependent
X--ray to optical spectral index $\alpha_{OX}$ \citep{cri03}. 
The correction for the number of missed AGNs in hard X--ray surveys
(i.e. Compton-thick) is obtained by requiring that the observed X--ray
luminosity function matches the X--ray background spectrum.  
The resulting estimate of $\rho_{\bullet,acc}$, assuming
$\epsilon$=0.1, is fully consistent with the local mass density 
$\rho_{\bullet,loc}$= 4.6 $^{+1.9}_{-1.4} \times 10^5 M_{\odot}$ Mpc$^{-3}$.
More recently, \citet{Ueda14} published an updated version of the
hard X--ray luminosity function, including an estimate of the Compton-thick population needed to fit the hard X--ray background spectrum. They
reached similar conclusions on the average accretion efficiency using
the \citet{Hop07} bolometric corrections and the \citet{Vika09} 
best fit value for  $\rho_{\bullet,loc}$= 4.9$\pm0.7 \times 10^5 M_{\odot}$ Mpc$^{-3}$.
\par
The agreement of the two approaches   described above  is not surprising 
despite the different assumptions of the  luminosity
functions and bolometric corrections. In fact, both  are linked to
reproduce the hard X--ray background, which can be considered an integral
constraint on the total mass accreted over the cosmic
time and locked in SMBH.
The almost identical values of the assumed local density
($\rho_{\bullet,loc}$)  thus lead to very similar conclusions on the
average accretion efficiency.

At  face value, and keeping all the factors entering in Eq. 1 
fixed at the fiducial values  described above, a higher value of the
integral local mass density would lead to a similarly lower value for
the average radiative  efficiency. The possibility that radiatively inefficient
accretion processes may be more common than previously thought has
been put forward by \citet{Novak13}. His argument is based on a simple
linear scaling of the commonly accepted value for the accretion
efficiency  ($\epsilon\sim$0.1) by a factor of 5, the maximum discrepancy 
between the revised estimate (\citealt{GS13}, \citealt{KO13}) and previous values.
Radiatively inefficient processes  (i.e. slim accretion disks) are
advocated to explain the fast growth of SMBH in the early Universe
(e.g. \citealt{Madau14}).  
 \par
The revised estimate of the local mass density of SMBH may have important consequences for AGN
demography  and accretion physics. In the following we discuss the
impact of an increased fraction of heavily obscured SMBH with respect
to current observational constraints.

\begin{figure*}
 \begin{center}
  \includegraphics[width=9.5truecm,height=13truecm,angle=-90]{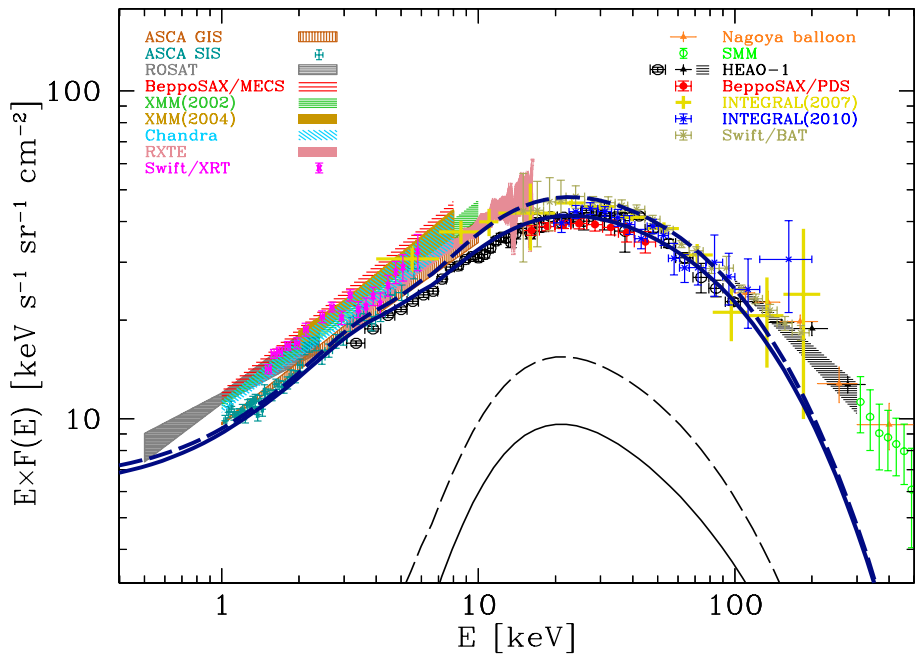}
  \caption{The broadband cosmic XRB spectrum. The thick, solid, dark blue curve
and the thin, solid, black curve show the total AGN spectrum and the
contribution of Compton-thick (mildly+heavily) AGNs in the GCH07 model,
respectively. The thick, dashed, dark blue curve and the thin, dashed,
black curve are as above, but assume a 4 times larger abundance of
heavily Compton-thick AGNs. The XRB measurements from various
mission/instrument combinations are marked by different symbols as
labeled. Labels on the left (right) generally refer to measurements at
$E<10$ ($>10$) keV. The relevant references are as follows. ASCA GIS:
Kushino et al. (2002); ASCA SIS: Gendreau et al. (1995); ROSAT:
Georgantopoulos et al. (1996); BeppoSAX/MECS: Vecchi et al. (1999);
XMM(2002): Lumb et al. (2002); XMM(2004): De Luca et al. (2004); RXTE:
Revnitsev et al. (2003); Swift/XRT: Moretti et al. (2009); Nagoya
balloon: Fukada et al. (1975); SMM: Watanabe et al. (1997); HEAO-1:
Gruber et al. (1999; open circles and filled triangles), Kinzer et
al. (1997; hatched area); BeppoSAX/PDS: Frontera et al. (2007);
INTEGRAL(2007): Churazov et al. (2007); INTEGRAL(2010): Turler et
al. (2010); Swift/Bat: Ajello et al. (2008).}
  \label{fig-xrb}
\end{center}
\end{figure*}

\section{A large population of deeply buried SMBH?}
\label{sec:bhmd}

The contribution to the SMBH mass density of unobscured and
obscured AGNs and the degeneracies intrinsic to the comparison between 
$\rho_{\bullet,acc}$ and $\rho_{\bullet,loc}$, are manifest if equation 1 is written as

\begin{equation}
  \rho_{\bullet,acc}c^2 
  = \langle k_{BOL} \rangle \frac{1- \epsilon}{\epsilon}  U_{XO} (1 +
  \sum R_{OBS}) 
\label{eq:grama3} 
,\end{equation}where $U_{XO}$ is the X--ray energy density of unobscured AGNs, as
determined from the soft X--ray luminosity function
(e.g. \citealt{grh05}), while $\langle k_{bol} \rangle$ is the average
bolometric correction which is assumed to be the same for obscured and
unobscured AGNs (e.g. \citealt{Beta12}). The sum is over the ratios to unobscured
AGNs of the obscured  population contributing to the SMBH mass density
and the X--ray background: 

\begin{equation}
  \sum R_{OBS} = R_{Thin} + R_{Thick} =  R_{Thin} + R_{MThick} + R_{HThick} \times (0.02/f_s) 
\label{eq:grama4} 
.\end{equation}

In the following we refer to the \citet{Marconi06} calculations, which are
based on the Gilli, Comastri \& Hasinger (2007,
GCH07)  AGN synthesis model where 
the population of Compton-thick (CT) AGNs is split into mildly
CT  ({\it MCT})  when the optical depth for Compton scattering $\tau_T$ is of the
order of a few, and heavily CT ({\it HCT}), or reflection-dominated, when  $\tau_T
\gg 1$. The former  provide the largest contribution to the XRB
(X--ray background) spectrum at 20--30
keV, while the contribution of the latter depends mainly on the
fraction of the reprocessed radiation $f_s$ reflected or scattered
towards the observer which, in turn, depends on the geometry of the absorber.
In GCH07 the $f_s$ value in the 2--10 keV band is assumed to be 0.02,
in broad agreement with the observations of Compton-thick AGNs in the
local Universe (see  Fig.~1 in GCH07). 
Moreover,  the ratio between  Compton-thin and unobscured
sources is luminosity dependent and smoothly decreases from $R=4$ at low luminosities
(log$L_X=42$)  to $R=1$ at high luminosities  (log$L_X=46$). The luminosity dependent obscured fraction is assumed to
be the same  for both Compton-thin and Compton-thick objects. As a
result, the total (thin plus thick)  obscured-to-unobscured ratio  decreases from 8:1 at low
luminosities to 2:1 at high luminosities. 
The luminosity averaged ratio is approximately 3 which means that for each unobscured AGN
there are three Compton-thin and three Compton-thick  sources (1.5 {\it
  MCT} and 1.5 {\it HCT}) and thus $\Sigma R_{OBS}$=6.
An excellent match between $\rho_{\bullet,acc}$ and
$\rho_{\bullet,loc}$, satisfying the XRB constraints, is found for
$\epsilon=$0.1 \citep{Marconi06}. 

We consider the most conservative case of a local mass density
which is a factor of 2 higher than  previously considered.
The revised value (\citealt{GS13}, \citealt{KO13}) is almost exactly a factor of
2 higher than that obtained by \citet{Marconi04} and \citet{Vika09}. 
We also assume that the additional mass density is not due 
to a lower efficiency, but  to a  new population  described by the
same  global parameters (cosmological evolution, bolometric correction, and
radiative efficiency)  of the known AGNs responsible for the XRB. 
If this were the case, equation 3 becomes
 
\begin{equation}
 2\rho_{\bullet,acc}c^2 = \langle k_{BOL} \rangle  U_{XO}  \frac{1- \epsilon}{\epsilon}  (1 +
 \sum_i R_{OBS} + R_{NEW}) 
\label{eq:grama5} 
,\end{equation}where $R_{NEW}$ is the  contribution of the new   population  responsible
for the local mass density excess.  It is straightforward to show
that 

\begin{equation}
 R_{NEW}  =  (1 + \sum R_{OBS}) = 7
\label{eq:grama6} 
,\end{equation}implying that the contribution to the mass density of the new population is the same as  that already
contributing to the XRB. 
It goes without saying that the new population would vastly exceed 
the limits imposed by the XRB itself, unless the sources are so
extremely obscured that X--rays are almost completely suppressed.
{\emph{In other words, the upward revised normalization can be matched if
each SMBH contributing to the XRB, has an X--ray quiet counterpart
that  adds mass, but almost no X--ray radiation.}}

This figure could be mitigated noting that the present uncertainties
on the XRB spectral intensity at $\simeq$ 20--30 keV are such that 
it would be possible to accommodate additional sources with the same
spectrum of {\it HCT} in GCH07.
More specifically, if the reflection-dominated sources are four times more numerous 
than in GCH07  (Fig.~1), the predicted XRB flux at
20--30 keV would still be consistent with the data, within the measurement errors. 
If this were the case the size of the  X--ray quiet population would
be  reduced to $R^{\arcmin}_{NEW}$ = 2.5.  This new value is derived by the  equality 

\begin{equation}
 R_{HCT}+ R_{NEW}  =  4 \times R_{HCT} + R^{\arcmin}_{NEW} 
\label{eq:grama7} 
;\end{equation}we note that  $R_{HCT}$ is 1.5. In this scenario, only about 18\% of the SMBH population would be
X--ray quiet, but having significantly increased the fraction of {\it
  HCT}  sources in the GCH07 model, the fraction of deeply buried
AGN ({\it HCT} plus X--ray quiet)  would be about 60\% of the total SMBH population. 

We can relax the GCH07 assumption of the  reflection yield
($f_s=0.02$), and assume that both the GCH07 model {\it HCT}
sources and the new population are described by a heavily
obscured, reflection-dominated spectrum, with the same value of 
the reflection yield in the 2--10 keV band. In this case the new
values of the yield of reflection-dominated sources are obtained from

\begin{equation}
 R_{HCT} \times 0.02   =  f^{\arcmin}_{NEW} \times [R_{HCT} + R_{NEW}] 
\label{eq:grama8} 
,\end{equation}
where $R_{HCT}$=1.5 is the fraction assumed in GCH07. With these
assumptions, we obtain  $f_{NEW}$=0.0035 and $f^{\arcmin}_{NEW}$=0.014
for $R_{NEW}$=7 or $R_{NEW}$=2.5, depending on the assumed level  of the XRB.

\section{Discussion}
\label{sec:discussion}

A large population of heavily obscured, reflection-dominated AGNs is
considered here to fill the mass gap between the most recent
estimate of $\rho_{\bullet,loc}$ and previous values. 
In order to accommodate a significantly higher  number of SMBH
without violating the limit imposed by the X--ray background, the
hypothetical new population should be characterized by an extremely
faint, almost vanishing, X--ray emission. 

The suppression of the hard X--ray emission  (up to a few hundred  keV)  would
require both extremely high column densities ($\tau \gg$ 10)
and large  covering factors (approaching $4\pi$).
The detection and the measure of column
densities in the high $\tau$ regime ($N_H \sim 10^{26}$ cm$^{-2}$
or even more) is extremely challenging. Large covering factors of the obscuring gas, with reflection yields
much lower than 1\%,  are invoked to explain the observations of a few 
heavily obscured and Compton-thick AGNs observed with {\it Suzaku} 
(e.g. \citealt{Ueda07}; \citealt{io10}). There is
also tentative evidence for a larger covering factor for increasing
absorption among Chandra Deep Field South faint sources \citep{bri14}.
The upper limits on the 2--10 keV X--ray emission of a small sample of
luminous AGNs, hosted by nearby ULIRGS  \citep{Nardini11}, are so tight
that the corresponding upper limits on $f_s$ are of the order of
$10^{-3} - 10^{-4}$.   If the presence of these AGNs is widespread
among infrared galaxies, they could be excellent candidates for the
new population. 

At face value,  a population of reflection-dominated Compton-thick AGNs,  with a suitable combination of covering factor and column
density,  would still be consistent with the upper envelope of  present uncertainties of the
hard XRB spectrum. 

While the new population could have so far escaped direct
detection in present deep, soft ($<$ 10 keV)  X--ray surveys, and
would remain largely undetected in shallow {\it NuSTAR} hard X--ray surveys
\citep{dave13}, the reprocessed AGN luminosity will emerge in the
infrared. 
A first order, integral estimate  may be obtained using the hard XRB spectral
intensity as a limit of the total energy output combined with suitable
bolometric corrections (e.g. \citealt{fi99}, \citealt{guido02}).
The results indicate that the AGNs responsible for the XRB make  10--15\%  of the IR
background around its $\sim$100$\mu$m peak. Their contribution
increases at shorter wavelengths, being of the order of  20--40\% at
60$\mu$m and possibly higher at 15$\mu$m depending on the  dust effective temperature  
\citep{guido02} and the relative fraction of Compton-thick sources in
the adopted XRB model. The impact of the new population of heavily
Compton-thick AGNs hypothesized here is even more uncertain.  
In the most naive approximation, the above estimates should be
rescaled by a factor of 2 making the AGN contribution  
dominant at  short wavelengths ($<$ 20 $\mu$m). 
However, according to a recent, comprehensive population synthesis model of the X--ray and infrared backgrounds 
\citep{Shi13}, the contribution of the X--ray emitting AGNs to the 100$\mu$m 
background is negligible and is of the order of 15\% at
5--10$\mu$m. If this were the case, there would be plenty of room to
fit a large population of  heavily obscured SMBH without exceeding the limits
imposed by the IR background.

Observational constraints on the abundance of sources likely to harbor a deeply
buried new AGN may be obtained by mid-infrared photometric and spectroscopic surveys 
carried out with the {\it Spitzer Space Telescope}.
Highly obscured and Compton-thick AGNs are likely to be affected by heavy dust absorption and steep
mid-infrared to optical slopes. A widely adopted selection method 
is based on the optical to 24$\mu$m  color  (e.g. \citealt{Fiore09}).
The optical depth of the silicate absorption feature at 9.7$\mu$m is 
also considered a reliable tracer of heavy dust absorption. 
Good quality, mid-infrared spectra were obtained for local luminous infrared galaxies in
the GOALS  survey \citep{stier13} and $z\sim$0.7 COSMOS galaxies
\citep{fu10}. 
The fraction of optically thick ($\tau_{9.7} >$1)  galaxies, which are
likely to host a buried AGN on the basis of SED decomposition, 
is estimated to be of the order of 10\% or even
lower. 
However, it should be noted that current mid-infrared photometric and spectroscopic
surveys  are relatively shallow and, moreover, 
there are many Compton-thick AGNs that do not display strong silicate
absorption features (e.g. \citealt{ioannis11}).
Given the present uncertainties  and the difficulty of distinguishing the
hypothetical population of deeply buried sources from normal and
star-forming galaxies, we conclude that the new
population,  is unlikely to exceed the IR background
spectral intensity.
 Candidates of the  new population may be found among sources
   with a steeply rising SED from the near- to the far-infrared 
  and/or among those with the highest extinction in the silicate
  absorption feature at 9.7$\mu$m.


\section{Conclusions}

We put forward the possibility that a sizeable
population of hitherto uncovered ultra obscured AGNs would explain the 
revised upward value of the SMBH local mass density $\rho_{\bullet,loc}$.

We are aware of the strong degeneracies between the various parameters
used to build our argument and, in particular, the radiative efficiency
$\epsilon$.  Within the reasonable assumptions
that the evolution of the X--ray luminosity function and
bolometric corrections are robustly constrained by recent surveys, and
the spectral intensity of the  X--ray background and the accretion processes
are self similar and not evolving, a large number of deeply buried
SMBH may be a viable solution beyond, radiatively inefficient accretion, for example.

This would imply that the column density distribution of nuclear obscuring
gas in SMBH is top heavy and skewed towards values of the order of $10^{25-26}$
cm$^{-2}$ or even higher and the covering fraction of obscuring
material may approach $4\pi$.
Finding observational proof of the existence of these ultra-obscured SMBH
is challenging. Deep  mid-far-infrared photometric and
spectroscopic surveys, matched with
sensitive X--ray surveys would be needed to estimate the fraction of 
infrared emitting AGNs without an X--ray counterpart and thus
presumably heavily buried. High density gas illuminated by a hidden
nuclear source may be traced by molecular transitions observable in
the far-IR with ALMA \citep{IN14}.
A fraction of them could be found among the 
sources responsible for the unresolved XRB in the 6--8 keV band 
\citep{Xue12}. We are hopeful that a few examples will be  uncovered by forthcoming  deep hard X--ray surveys with {\it NuSTAR} 
\citep{fiona13} and ultra-deep ATHENA surveys.



\begin{acknowledgements}
This work was partially supported by the  ASI/INAF  I/037/12/0--011/13,
the PRIN--INAF--2011 and the PRIN--INAF--2012 grants. We thank Marcella Brusa, Marco
Mignoli and Gianni Zamorani for useful discussions, and the anonymous
referee for constructive comments. AC acknowledges the support of a 
Caltech Kingsley visitor fellowship.
\end{acknowledgements}


\begin{thebibliography}{}

\bibitem[Ajello et al. (2008)]{ajello08}
Ajello,~M., Greiner, J., Sato, G., et al. 2008, 
\apj, 689, 666

\bibitem[Alexander et al. (2013)]{dave13}
Alexander, D.M.; Stern, D., Del Moro, A., et al. 2013,
\apj, 773, 125

\bibitem[Boyle et al. (2000)]{Boy00}
Boyle,~B.J., Shanks,~T., Croom,~S.M., et al. 2000,
\mnras, 317, 1014

\bibitem[Brightman et al. (2014)]{bri14}
Brightman, M., Nandra, K., Salvato, M., et al. 2014,
\mnras, 443, 1999

\bibitem[Churazov et al. (2007)]{churazov07}
Churazov, E., Sunyaev, R., Revnivtsev, M., et al. 2007,
\aap, 467, 529

\bibitem[Comastri et al. (2010)]{io10}
Comastri, A., Iwasawa, K., Gilli, R., et al. 2010,
\apj, 717, 787

\bibitem[De Luca \& Molendi (2004)]{deluca04}
De Luca,~A., Molendi,~S. 2004,
\aap, 419, 837

\bibitem[Elvis et al. (2002)]{Elvis02}
Elvis~M., Risaliti,~G., Zamorani,~G. 2002,
\apj, 565, L75

\bibitem[Fabian \& Iwasawa (1999)]{fi99}
Fabian~A.C., Iwasawa,~K. 1999,
\mnras, 303, L34

\bibitem[Fiore et al. (2009)]{Fiore09}
Fiore~F., et al. 2009,
\apj, 693, 447

\bibitem[Frontera et al. (2007)]{ff07}
Frontera,~F., Orlandini, M., Landi, R., et al. 2007, 
\apj, 666, 86

\bibitem[Fu et al. (2010)]{fu10}
Fu,~H., Lin,~Y., Scoville,~N.Z., et al. 2010,
\apj, 722, 653

\bibitem[Fukada et al. (1975)]{fukada75}
Fukada, Y., Hayakawa, S., Kasahara, I., et al. 1975, 
\nat, 254, 398

\bibitem[Gendreau et al. (1995)]{gendreau95}
Gendreau, K.C., Mushotzky, R.F., Fabian, A.C., et al. 1995, 
\pasj, 47, L5

\bibitem[Georgantopoulos et al. (1996)]{ioannis96}
Georgantopoulos, I., Stewart, G.C., Shanks, et al. 1996, 
\mnras, 280, 276

\bibitem[Georgantopoulos et al. (2011)]{ioannis11}
Georgantopoulos,~I., et al. 2011, 
\aap, 531, A116

\bibitem[Gilli et al. (2007)]{gilli2007}
Gilli,~R., Comastri,~A., \& Hasinger,~G. 2007, 
\aap, 463, 79   GCH07 

\bibitem[Graham \& Scott (2013)]{GS13}
Graham,~A.W., Scott,~N. 2013, 
\apj, 764, 151

\bibitem[Gruber et al. (1999)]{gruber99}
Gruber, D.E., Matteson, J.L., Peterson, L.E., Jung, G.V. 1999, 
\apj, 520, 124

\bibitem[Harrison et al. (2013)]{fiona13}
Harrison, F.A., et al. 2013,
\apj, 770, 103

\bibitem[Hasinger et al. (2005)]{grh05}
Hasinger, G., Miyaji, T., Schmidt, M. 2005,
\aap, 441, 417 

\bibitem[Hopkins et al. (2007)]{Hop07}
Hopkins, P.F., Richards, G.T., Hernquist, L. 2007,
\apj, 654, 731

\bibitem[Imanishi \& Nakanishi (2014)]{IN14}
Imanishi~M., Nakanishi,~K. 2014,
\aj, 148, 9

\bibitem[Kinzer et al. (1997)]{kinzer97}
Kinzer, R.L., Jung, G.V., Gruber, D.E., et al. 1997,
\apj, 467, 361

\bibitem[Kormendy \& Ho (2013)]{KO13}
Kormendy,~J., Ho,~L.C. 2013,
\araa, 51, 511

\bibitem[Kushino et al. (2002)]{kushino02}
Kushino, A., Ishisaki, Y., Morita, U., et al. 2002, 
\pasj, 54, 327

\bibitem[Lumb et al. (2002)]{lumb02}
Lumb, D.H., Warwick, R.S., Page, M., De Luca, A. 2002, 
\aap, 389, 93

\bibitem[Lusso et al. (2012)]{Beta12}
Lusso,~E., Comastri,~A., Simmons,~B.D., et al. 2012,
\mnras, 425, 623 

\bibitem[Madau et al. (2014)]{Madau14}
Madau,~P., Haardt,~F., Dotti,~M. 2014,
\apj, 784, L38

\bibitem[Marconi \& Hunt (2003)]{marconi03}
Marconi,~A., Hunt~L.K. 2003,
\apj, 589, L21

\bibitem[Marconi et al. (2004)]{Marconi04}
Marconi,~A., Risaliti,~G., Gilli,~R., et al. 2004,
\mnras, 351, 169

\bibitem[Marconi et al. (2006)]{Marconi06}
Marconi,~A., Comastri,~A., Gilli,~R., et al. 2006,
Mem. S.A.It. 77, 742

\bibitem[McLure \& Dunlop (2002)]{mclure02}
McLure,~R. J., Dunlop,~J. S. 2002,
\mnras, 331, 795

\bibitem[Merloni \& Heinz (2008)]{merloni08}
Merloni, A., Heinz,~S. 2008,
\mnras, 388, 1011

\bibitem[Merrit \& Ferrarese (2001)]{mf01}
Merrit,~D., Ferrarese,~L. 2001,
\apj, 547, 140

\bibitem[Miyaji et al. (2000)]{taka00}
Miyaji,~T. Hasinger,~G., Schmidt,~M. 2000,
\aap, 353, 25 

\bibitem[Nardini \& Risaliti (2011)]{Nardini11}
Nardini,~E., Risaliti,~G. 2011,
\mnras,  415, 619

\bibitem[Novak (2013)]{Novak13}
Novak, G.S. 2013,
\mnras, submitted, arXiv:1310.3833

\bibitem[Revnivtsev et al. (2005)]{revni05}
Revnivtsev,~M., Gilfanov,~M., Jahoda,~K., Sunyaev, R. 2005,
\aap, 444, 381

\bibitem[Risaliti et al. (2002)]{guido02}
Risaliti, G., Elvis, M., Gilli, R. 2002, 
\apj, 566, L67 

\bibitem[Sani et al. (2011)]{Sani11}
Sani,~E., Marconi,~A., Hunt,~L.K., Risaliti,~G. 2011,
\mnras, 413, 1479

\bibitem[Shakura \& Sunyaev (1973)]{SS73}
Shakura,~N.I., Sunyaev,~R.A. 1973,
\aap, 24, 337

\bibitem[Shankar et al. (2004)]{shankar04}
Shankar,~F., Salucci,~P., Granato,~G.L., et al. 2004,
\mnras, 354, 1020

\bibitem[Shi et al. (2013)]{Shi13}
Shi,~Y., Helou,~G., Armus,~L., et al. 2013,
\apj, 764, 28

\bibitem[Soltan (1982)]{Sol82}
Soltan,~A. 1982,
\mnras, 200, 115

\bibitem[Stierwalt et al. (2013)]{stier13}
Stierwalt,~S., Armus,~L., Surace,~J.A., et al. 2013, 
\apjs, 206, 1


\bibitem[Turler et al. (2010)]{turler10}
Turler,~M., Chernyakova, M., Courvoisier, T.J.L., et al. 2010, 
\aap, 512, A49

\bibitem[Ueda et al. (2003)]{Ueda03}
Ueda,~Y. Akiyama,~M., Ohta,~K., Miyaji,~T. 2003,
\apj, 598,886

\bibitem[Ueda et al. (2007)]{Ueda07}
Ueda, Y., Eguchi, S., Terashima, Y., et al. 2007, 
\apj, 664, L79

\bibitem[Ueda et al. (2014)]{Ueda14}
Ueda, Y., Akiyama, M., Hasinger, G., et al. 2014,
\apj, 786, 104

\bibitem[Vecchi et al. (1999)]{vecchi99}
Vecchi,~A., Molendi,~S., Guainazzi,~M., et al. 1999, 
\aap, 349, L73

\bibitem[Vignali et al. (2003)]{cri03}
Vignali,~C., Brandt,~W.N., Schneider,~D.P. 2003,        
\aj, 125, 433   

\bibitem[Vika et al. (2009)]{Vika09}
Vika, M., Driver, S.P., Graham,~A.W., Liske,~J. 2009, 
\mnras, 400, 1451 

\bibitem[Watanabe et al. (1997)]{watanabe97}
Watanabe,~K., et al. 1997, 
in The Fourth CGRO Symposium, ed. C. D. Dermer, M. S. Strickman, \& J. D. Durfess, 
AIP Conf. Proc., 410, 1223

\bibitem[Xue et al. (2012)]{Xue12}      
Xue, Y.Q., Wang, S.X., Brandt, W.N., et al. 2012,
\apj, 758, 129 

\end{thebibliography}
\end{document}